\documentclass[preprint]{elsarticle} 
\usepackage{graphicx}
\usepackage{color}
\usepackage{amsmath}
\usepackage{amssymb}
\biboptions{numbers,sort&compress}
\newcommand{\beq}{\begin{equation}}
\newcommand{\eeq}{\end{equation}}
\newcommand{\bea}{\begin{eqnarray}}
\newcommand{\eea}{\end{eqnarray}}

\newcommand{\finalresult}{\kappa = 0.0149 \pm 0.0021}
\begin{document} 
\begin{frontmatter}

\title{The QCD phase diagram from analytic continuation}

\author{
R. Bellwied$^{5}$, 
S. Borsanyi$^{1}$, 
Z. Fodor$^{1,2,3}$, 
J. G\"unther$^{1}$, 
S. D. Katz$^{2,4}$,\\ 
C. Ratti$^{5}$,
K.K. Szabo$^{1,3}$
\\
$^1$ {\it Department of Physics, Wuppertal University, Gaussstr. 20, D-42119 Wuppertal, Germany}\\
$^2$ {\it Inst. for Theoretical Physics, E\"otv\"os University,}\\
{\it P\'azm\'any P. s\'et\'any 1/A, H-1117 Budapest, Hungary}\\
$^3$ {\it J\"ulich Supercomputing Centre, Forschungszentrum J\"ulich, D-52425
J\"ulich, Germany}\\
$^4$ {\it MTA-ELTE "Lend\"ulet" Lattice Gauge Theory Research Group,}\\
{\it P\'azm\'any P. s\'et\'any 1/A, H-1117 Budapest, Hungary}\\
$^5$ {\it Department of Physics, University of Houston, Houston, TX 77204, USA}
}

\date{\today}

\begin{abstract}
We present the crossover line between the quark gluon plasma and the
hadron gas phases for small real chemical potentials.
First we determine the effect of imaginary values of the chemical potential on
the transition temperature using lattice QCD simulations.  Then we use various
formulas to perform an analytic continuation to real values of the
baryo-chemical potential. Our data set maintains strangeness neutrality
to match the conditions of heavy ion physics.
The systematic errors are under control up to $\mu_B\approx 300$ MeV.
For the curvature of the transition line we find that there is an approximate
agreement between values from three different observables:
the chiral susceptibility, chiral condensate and strange quark susceptibility.
The continuum extrapolation is based on $N_t=$ 10, 12 and 16 lattices.
By combining the analysis for these three observables we find, for the curvature, the value $\finalresult$.
\end{abstract}

\end{frontmatter}

\section{Introduction\label{intro}}

For heavy ion physics, the most important feature of the phase diagram
of Quantum Chromodynamics (QCD) is the line that separates the hadron
gas phase from the quark gluon plasma, and the conjectured critical end-point
along this line separating cross-over from first order transition
\cite{Stephanov:2007fk}.

The qualitative form of the phase diagram was sketched four decades ago
\cite{Cabibbo:1975ig} as a consequence of Hagedorn's exponential spectrum of
hadron masses \cite{Hagedorn:1965st}.  The order of the transition at zero
density has been determined much later, for Nature's selection of quark masses
the two high temperature phases are connected through a cross-over
\cite{Aoki:2006we}. In the absence of a real transition the cross-over
temperature can be determined but it is ambiguous \cite{Aoki:2006br}.
Observables that are related to the spontaneous breaking of chiral symmetry
(chiral condensate and its susceptibility) give a temperature around 155~MeV
\cite{Aoki:2006br,Aoki:2009sc,Borsanyi:2010bp,Bazavov:2011nk}.

Beyond the transition temperature $T_c$ at vanishing density, 
the chiral cross-over line is described by a standard
curvature parameter ($\kappa$) and higher order terms:
\begin{equation}
\frac{T_c(\mu_B)}{ T_c(\mu=0)}=1 
- \kappa \left(\frac{\mu_B}{T_c(\mu_B)}\right)^2
+ \lambda \left(\frac{\mu_B}{T_c(\mu_B)}\right)^4\dots
\label{eq:kappa}
\end{equation}

Extracting $T_c(\mu_B)$ from first principles is very challenging. Direct 
Monte-Carlo calculations are hindered by the sign problem.
Attempts to reach non-vanishing $\mu_B$ on the lattice include 
reweighting of the generated configurations \cite{Barbour:1997ej,Fodor:2001au,Fodor:2001pe,Csikor:2004ik,Fodor:2004nz}, Taylor expansion in $\mu$ \cite{Allton:2002zi,Allton:2005gk,Gavai:2008zr,Basak:2009uv,Kaczmarek:2011zz}, analytic continuation from imaginary $\mu$ \cite{deForcrand:2002ci,DElia:2002gd,Wu:2006su,DElia:2007ke,Conradi:2007be,deForcrand:2008vr,DElia:2009tm,Moscicki:2009id}, use of the canonical ensemble \cite{Alexandru:2005ix,Kratochvila:2005mk,Ejiri:2008xt} and density of state methods \cite{Fodor:2007vv,Alexandru:2014hga}.  More recent approaches are represented by the use of dual variables
\cite{Gattringer:2014nxa}, and the complex Langevin equation
\cite{Seiler:2012wz,Sexty:2013ica}. However, their application to QCD with
physical parameters and controlled discretization has not yet been achieved.
The phase diagram was frequently studied in various model frameworks,
see e.g. Ref.~\cite{Stephanov:2007fk} and references therein. 
Recently, functional methods have also been applied to QCD~\cite{Fischer:2012vc,Fischer:2013eca,Fischer:2014ata}.

For the first few coefficients in Eq.~(\ref{eq:kappa}) it is enough to
study QCD at small $\mu_B$, for which there are several methods.
$\kappa$ can be and has been determined by calculating the $\mu_B$-derivative of
the chiral condensate using only $\mu_B=0$ ensembles
\cite{Kaczmarek:2011zz,Endrodi:2011gv}. 
However, the signal/noise ratio of higher $\mu_B$ derivatives is suppressed
with powers of the volume, making this approach impractical beyond $\mu_B^2$
order. Lattice calculations are perfectly feasible, though, with imaginary values of
the chemical potential \cite{Fodor:2001au,deForcrand:2002ci,Philipsen:2007rj}.
Setting $\mu_B=i\mu^I_B$ one avoids the sign problem
and the transition line can be studied
\cite{Cea:2006yd,Cea:2010md,Cea:2014xva,Bonati:2014rfa}.

In this study we follow the imaginary-$\mu_B$ approach and
go beyond previous studies by
a) performing a continuum approximation with lattices up to $N_t=16$; 
b) tuning $\mu_S(\mu_B,T)$ such that the strangeness
neutrality condition is maintained; c) using several observables: chiral
condensate, chiral susceptibility and strange susceptibility;
d) comparing the Taylor and the imaginary-$\mu$ method for
the strange susceptibility;
e) calculating the systematic errors
from scale setting, fit ranges, analytic formulas, etc.\footnote{During the writing
of this manuscript a similar independent analysis, based on $N_t=6,8,10,12$ lattices
and analytic continuation from imaginary $\mu_B$ appeared in arXiv~\cite{Bonati:2015bha}. Their
findings are similar to ours but the present analysis has
finer lattices, smaller pion splittings and significantly
larger statistics.
}

This Letter first gives a brief account of the necessary lattice 
simulations at zero and finite temperatures. Then the method for
setting strangeness neutrality is explained. Finally, we give a detailed description
of the analysis and present the continuum results for the curvature.

\section{Simulation setup}

This study is part of the 2nd generation staggered thermodynamics program of
the Wuppertal-Budapest collaboration \cite{Bellwied:2015lba}. We use a four times stout
\cite{Morningstar:2003gk} smeared ($\rho=0.125$) staggered fermion action with
2+1+1 flavors, i.e. dynamical up, down, strange and charm quarks. The gauge action uses the
tree-level Symanzik improvement.  The two light quarks are degenerate, their
masses and the strange quark mass are tuned such that the physical pion 
and kaon mass over pion decay constant are reproduced for
every lattice spacing.
For the zero temperature runs we kept the volume large $Lm_\pi>4$ in the
entire lattice spacing range of interest for this study: $a=0.2 \dots 0.063$~fm.
The charm mass was set to $m_c/m_s=11.85$ \cite{McNeile:2010ji}. 
The simulation parameters are detailed in Ref~\cite{Bellwied:2015lba}.
The overall scale was determined from $f_\pi$. We used $w_0$
as an alternative scale setting for the analysis~\cite{Borsanyi:2012zs}.

The chiral susceptibility as well as the chiral condensate require
renormalization. The additive divergence is removed by subtracting
the vacuum expectation value, the multiplicative divergence canceled
by the same factor in the bare quark mass \cite{Aoki:2006br,Aoki:2009sc}.
The renormalized condensate and its susceptibility are dimensionful
quantities, we use the fourth power of the pion mass to form a dimensionless
observable. We do not restrict the chiral susceptibility to the disconnected
part. The third observable that we use to identify the $\mu$-dependent
transition temperature is the strange susceptibility: thanks to the exact
quark number conservation it does not require renormalization.

At finite temperature, we have collected data at zero and at imaginary
baryo-chemical potentials. The $\mu_B=0$ data are used to perform a Taylor
expansion on one of our studied observables, and also to obtain a ``baseline''
for the shifted transition temperatures at other $\mu_B$ values. The
zero density configurations are listed in Ref.~\cite{Bellwied:2015lba}. 

The range of imaginary baryo-chemical potentials is limited by the Roberge-Weiss
transition at $\mu_B=i\pi T$ \cite{Roberge:1986mm}.  Below a limiting temperature $T_{RW}$ there is no transition as  $\mathrm{Im}\left[\mu_B/T\right]$ crosses $\pi$, but there is a first order transition above $T_{RW}$, where the imaginary density is non-vanishing and flips sign at $\mu^I_B/T=\pi$. The nature of the transition at
$T_{RW}$ depends on the quark masses
\cite{DElia:2009qz,Philipsen:2014rpa,Wu:2014lsa}. For intermediate masses the
system at $T=T_{RW}$ and $\mu^I_B/T=\pi$ will be critical, and then
in the entire range of smaller imaginary chemical potentials we will
see a crossover in temperature. Our data suggests that, for physical
masses, the latter scenario is realized, namely we are working with
a cross-over for all used $\mu_B^I/T$.

We selected six imaginary chemical potential values:
\begin{equation}
\mu_B^{(j)}= i T \frac {j\pi}{8}\,,\quad j=1,2,3,4,5,6
\label{eq:muchoice}
\end{equation}
We have all six $j$ values for our $N_t=$8, 10 and 12 lattices and
only $j=3 \dots 6$ for $N_t=$16. The reason for this is the following:
$j=0\dots 5$ data are
needed to determine the simulation parameters at finite
imaginary $\mu_B^{(j+1)}$ such that the strangeness neutrality condition
is fulfilled (see later).
The continuum extrapolation for this analysis can be carried out using $N_t=$8, 10 and 12
lattices. For the determination of the curvature $\kappa$ of the 
phase diagram we also need the finest $N_t=16$ lattices. Since $j=$1 and 2
do not give a statistically very significant contribution to $\kappa$ we
decided not to have these two points in our most expensive $N_t=16$ ensembles.
Therefore, in order to have the same setup for all lattice spacings, 
the $\kappa$ determination is based on $j=3,4$ and $5$. 
The $j=6$ point is used to estimate higher order effects.

This range to find the $\kappa$ coefficient ($\mu^I_B/T\lesssim2$) is narrower
than in earlier studies (e.g.  $\mu^I_B/T \lesssim 2.36$ in
Ref.~\cite{Cea:2014xva} and $\mu^I_B/T\lesssim
2.6$ in Ref.~\cite{Bonati:2014rfa}). A broader range of chemical potentials
has the advantage that the numerical derivative $[T_c(\mu_B)-T_c(0)]/\mu_B^2$
has a larger signal/noise ratio. However, more
non-linearities appear in a broader range and the results are more prone
to systematic errors as the singularity at $\mu^I_B\approx\pi T$ is approached.
This is the reason (to avoid unwanted systematic uncertainties) why we have taken a 
smaller $\mu^I_B$ range and we use other methods to increase the signal/noise ratio.

We performed simulations on $32^3\times 8$, $40^3\times10$, $48^3\times12$ and $64^3\times16$ lattices, at sixteen temperatures in the temperature
range 135\dots 210 MeV.
We have generated between 10000-15000 Hybrid Monte Carlo updates,
analyzing every 5th of them (every 10th for $N_t=16$). The configurations have
been evaluated for up to fourth order generalized quark number
susceptibilities \cite{Borsanyi:2013hza} and for the chiral condensate
and susceptibility. For $\mu_B=0$ we have $5\dots10$ times more statistics,
this ensures a solid guidance to the fitting procedure.

\section{Strangeness neutrality}

The most popular representation of the QCD phase diagram is in the temperature
vs. chemical potential plane. The baryo-chemical potential axis
leaves room for various interpretations. Ref.~\cite{Cea:2014xva} used the full
baryo-chemical potential including the strange quarks, i.e.
$\mu_u=\mu_d=\mu_s=\mu_B/3$. In Ref.~\cite{Bonati:2014rfa} 
both $\mu_u=\mu_d=\mu_s=\mu_B/3$ and $\mu_u=\mu_d=\mu_B/3$, $\mu_s=0$ were studied. 

However, neither of the recipes $\mu_s=0$ or $\mu_s=\mu_B/3$ maps consistently to the
situation that is realized in experiment. In heavy ion collisions, non-strange
particles are colliding. Although $s{\bar s}$ are generated in the collision, the net-strangeness is zero. 
Therefore we want to tune the chemical potentials to such 
values which guarantee strangeness neutrality. The light 
chemical potentials are kept identical ($\mu_u=\mu_d$), which ensures
isospin symmetry also at finite $\mu_B$. This corresponds to an experimental
situation where $Z=0.5 A$. Alternatively, one can achieve a different $Z/A$ ratio
corresponding to heavy nuclei by tuning $\mu_u$ and $\mu_d$ appropriately. This
possibility will be discussed later. Requiring strangeness neutrality
and fixing the value of $Z/A$ (or alternatively the electric charge/baryon number
ratio) uniquely determines all three quark chemical potentials as
functions of $\mu_B$. For the isospin symmetric case $\mu_u=\mu_d=\mu_B/3$ 
so the only non-trivial task is to find the strange quark or strangeness
chemical potential.

The strange quark chemical potential ($\mu_s$) is related to the strangeness ($\mu_S$) and 
baryo-chemical potential ($\mu_B$) as $\mu_s=\mu_B/3-\mu_S$. Then $\mu_s=\mu_B/3$ 
approximates strangeness neutrality at low temperature, and $\mu_s=0$ 
at high temperature. In this work we determine the strangeness
neutral trajectory $\mu_S(\mu_B,T)$ from lattice simulations. 

It is relatively straightforward to perform Taylor expansions from $\mu_B=0$ on
the trajectory that respects strangeness neutrality. For the equation of state
\cite{Borsanyi:2012cr,Hegde:2014wga} and for fluctuations relevant for
calculating freeze-out parameters in heavy ion collisions
\cite{Bazavov:2012vg,Borsanyi:2013hza} this procedure is already standard.

\begin{figure}
\begin{center}
\includegraphics[width=3in]{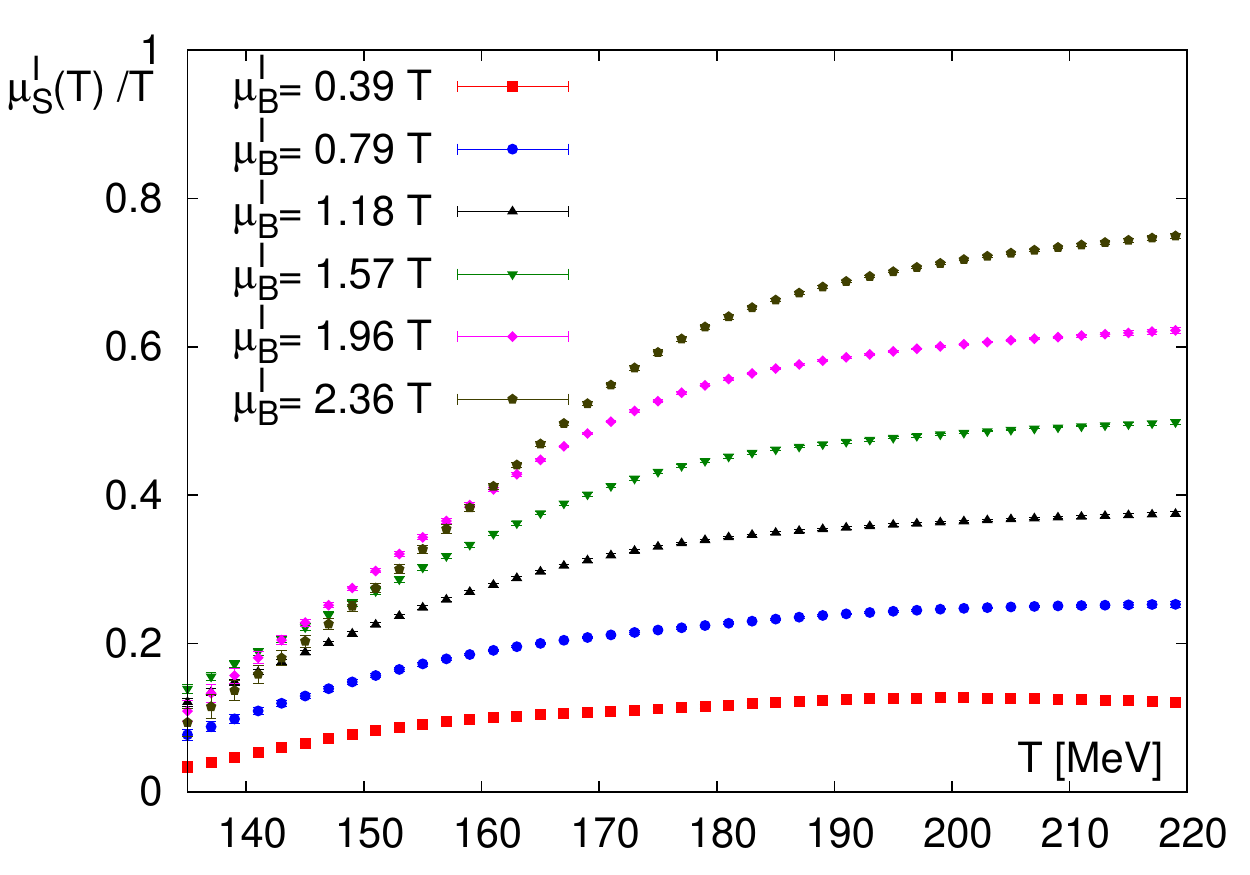}
\end{center}
\caption{\label{fig:mus}
The imaginary strangeness chemical potential that realizes
strangeness neutrality. Here we show a continuum extrapolation
based on $32^3\times8$, $40^3\times10$, $48^3\times 12$ lattices for
$\mu_B^I/T=0.39$ and $0.79$ and also using $64^3\times 16$ for the
larger chemical potentials. The error is statistical only.
}
\end{figure}

For actual simulations at finite $\mu^I_B$ the strange chemical potential
has to be fine tuned for every temperature, baryo-chemical potential
and lattice spacing. We solved this challenge by solving the
\begin{equation}
\frac{d}{d\mu_B^I} \frac{\partial \log Z}{\partial \mu_S} =0\,,
\end{equation}
differential equation discretized in $\mu_B$ with the trivial initial condition 
$ \partial \log Z/ \partial \mu_S =0$ at $\mu_B^I=0$. This equation simply states
that the $\mu_B^I$ derivative of strangeness is zero.
Using the 2nd order explicit
Runge-Kutta scheme,
we determine $\mu^I_S(\mu^I_B)$ using the prescription:
\begin{equation}
\mu^I_S(\mu^I_B+\Delta\mu^I_B) = \mu^I_S(\mu^I_B-\Delta\mu^I_B)
-2 \left.\frac{\chi_{SB}^{11}}{\chi_S^2}\right|_{\mu^I_B} \Delta\mu^I_B\,,
\label{eq:rungekutta}
\end{equation}
with the step size $\Delta \mu^I_B/T = \pi/8$ (see Eq.~\ref{eq:muchoice}).
For the initial step ($\mu_B^I/T=\Delta \mu^I_B/T$) we used the high-statistics
$\mu_B=0$ runs and NLO Taylor expansion.
Each step using Eq.~(\ref{eq:rungekutta}) requires a simulation
at $\mu_B^I$ and the evaluation of the 2nd order fluctuations:
$\chi_{SB}^{11}=1/(TV) \partial^2 \log Z/\partial\mu_S\partial \mu_B$ and
$\chi^S_2 = 1/(TV) \partial^2 \log Z/\partial \mu_S^2$.
This method would be $\mathcal{O}({\Delta\mu^I_B}^2)$
accurate only, but as an additional correction, we do a small extrapolation
for both terms on the RHS of Eq.~(\ref{eq:rungekutta})
after each simulation so that the remaining strangeness neutrality
violation is not propagated to the next step.  For this extrapolation,
we need higher order fluctuations \cite{Borsanyi:2013hza}.
This combination is $\mathcal{O}({\Delta\mu^I_B}^3)$ accurate
in the complete $\mu_B^I$ range. The resulting $\mu_S(\mu_B,T)$ 
function is interpolated in $T$ and extrapolated in $1/N_t^2$
and the resulting smooth function is used to start the simulations at 
$\mu_B^I+\Delta\mu_B^I$.
In Fig.~\ref{fig:mus} we show the resulting strangeness chemical potential.

\begin{figure}[t]
\begin{center}
\includegraphics[width=3in]{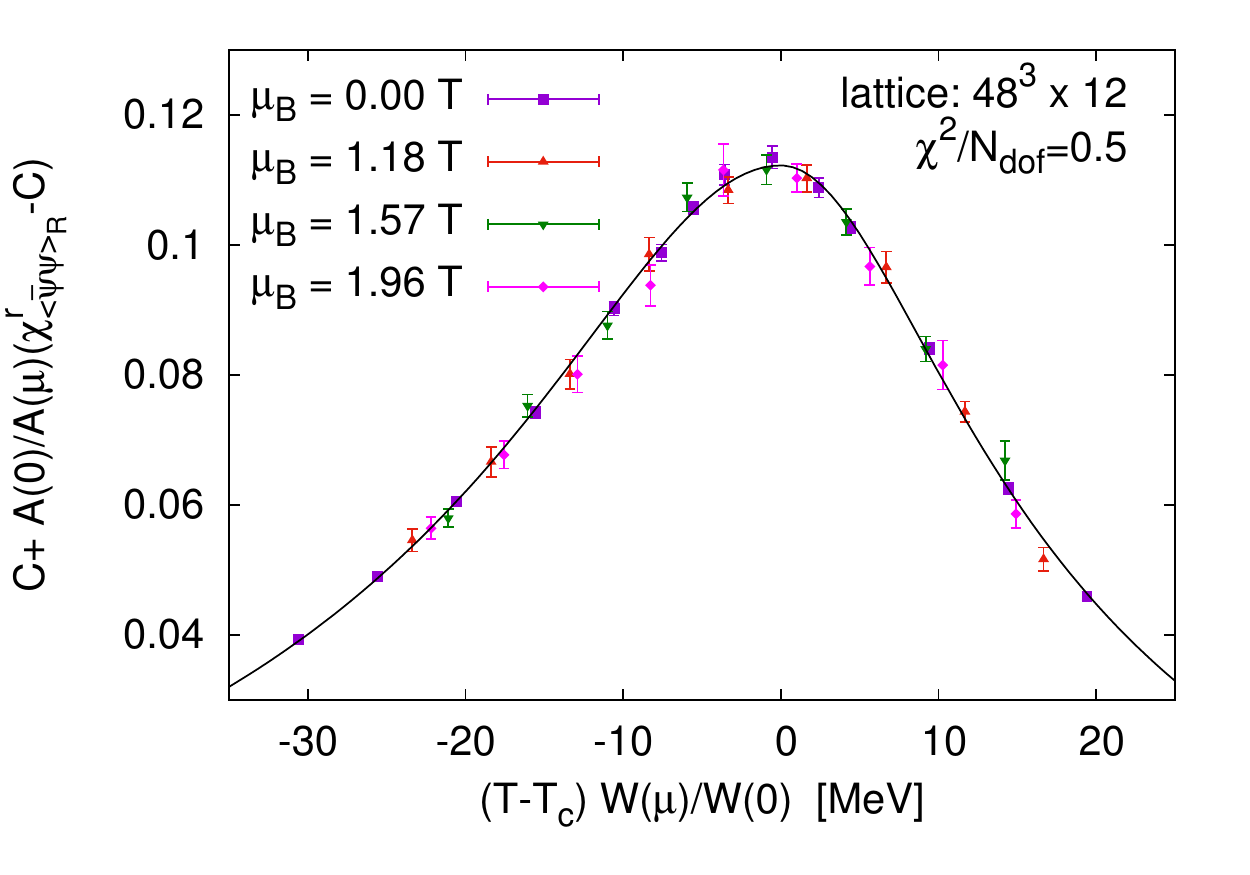}
\end{center}
\caption{\label{fig:pbpsusc}
The chiral susceptibility at several imaginary chemical potentials
on our $48^3\times12$ lattice.  After a $\mu_B^I$-dependent shifting and
stretching, data from all chemical potentials collapse on one curve.
The fitted curve corresponds to eqn.(\ref{eq:chi_fit1}) at $\mu_B=0$.
}
\end{figure}

\section{Analysis details}

We calculate the curvature of the phase diagram from three observables.
We calculate statistical and systematic errors for all three. 

1) Our first observable is the chiral susceptibility
$\chi_{\bar\psi\psi}/m_\pi^4$. As discussed previously, it requires additive and
multiplicative renormalization. For details on this procedure see 
Ref.~\cite{Borsanyi:2010bp}.  The chiral susceptibility forms a peak at the transition temperature. With
increasing imaginary chemical potential this peak is shifted towards higher
temperatures, approximately maintaining its height and width. For other
normalizations (e.g. $\chi_{\bar\psi\psi}/T^4$) the shape of the function
changes more significantly while varying the chemical potential. 

We fit $\chi_{\bar\psi\psi}(\mu_B^I,T)/m_\pi^4$ in a global fit function
where for each $\mu_B$ a different width, height and peak position
is allowed, but the other parameters that describe the peak shape
are $\mu_B$-independent. We use two different modifications to the Lorentzian
peak form: 
\beq
\frac{\chi_{\bar\psi\psi}^r (\mu,T)}{m_\pi^4}= \begin{cases}
  C + A^2(\mu) \left( 1+W^2(\mu)(T-T_c(\mu))^2 \right)^{\alpha} \ \ \ \text{for} \  T \leq T_c \\
   C + A^2(\mu) \left( 1+B^2W^2(\mu)(T-T_c(\mu))^2 \right)^{\alpha} \ \ \ \text{for} \ T > T_c
 \end{cases} \label{eq:chi_fit1}
\eeq
and
\beq
\frac{\chi_{\bar\psi\psi}^r(\mu,T)}{m_\pi^4} = C + \frac{A(\mu)}{1+W^2(\mu) (T-T_c(\mu))^2 + BW^3(\mu) (T-T_c(\mu))^3} 
\eeq
The $\mu$ dependent parameters $A(\mu)$, $W(\mu)$ and $T_c(\mu)$ describe
the change in the height, width and the position of the curve 
as $\mu$ increases.
For the zero temperature data which are required for renormalization we
use two different interpolations in the inverse gauge coupling: a 6th 
order polynomial and a simple rational function.
We have two options for the scale setting using
$f_\pi$ or $w_0$ and we apply three possible fit windows to
select the transition range. In order not to interfere with the shifted
temperature dependence the fit windows constrain the value of the
susceptibility, not the temperature.

The effect of the $\mu$ dependent parameters is a shift in $T$, and 
a rescaling in $T$ and $\chi$. Applying the inverse transformation
to the finite $\mu_B^I$ data points all of them should collapse on
the $\mu_B=0$ curve. This is demonstrated in Fig.~\ref{fig:pbpsusc}. 
The advantage of this procedure is that the $\mu$ independent parameters
can be fitted using the the high-statistics runs at $\mu_B=0$ and the
non-vanishing $\mu_B^I$ runs are needed to determine the relative
position and rescaling compared to this more complicated functional form. 
This allows the precise determination of $\Delta T_c(\mu_B^I)$ with an error
below 0.25 MeV, while $T_c(\mu_B)$ itself has an error of several MeV.
We extract $\kappa$ by a linear fit of $\Delta T_c$ vs. $\mu_B^2$
in the range $1.2 \lesssim \mu_B^I/T \lesssim 2$, and extrapolate $\kappa$ to the continuum.
Since the continuum extrapolation of $\kappa$ had a large $\chi^2$
when all four lattices were used we included only $N_t=10, 12$ and 16 in
the final result, resulting in a good $\chi^2$ for all analyses.
In an alternative analysis we made a combined continuum and $\mu_B^2$ fit again
using only the finest three lattice spacing, and found acceptable $\chi^2$ values again.

2) The chiral condensate $\langle\bar\psi\psi\rangle^r= m_q (d\log
Z/dm_q)/m_\pi^4$ is a remnant order parameter of the chiral transition.
Its inflection point (though it is hard to locate in a finite precision
data set) is very close to the peak position of $\chi_{\bar\psi\psi}/m_\pi^4$.
At finite $\mu_B^I$  the temperature dependence of $\langle\bar\psi\psi\rangle^r$
is shifted and very slightly stretched.

We find that the data at $\mu_B=0$ (see Ref.~\cite{Borsanyi:2010bp}) can be very accurately described by
simple fit functions. The $\mu$ dependence in this case is well described
by just two $\mu_B$-dependent parameters describing a shifting and rescaling of 
the renormalized condensate. We use the following parameterizations:
\beq
\langle\bar\psi\psi\rangle^r(\mu,T) = A(\mu) \left( 1+ B \tanh\left[ C \left( T-T_c(\mu) \right)  \right] + D \left( T-T_c(\mu) \right)
\right)
\eeq
and
\beq
\langle\bar\psi\psi\rangle^r(\mu,T) = A(\mu) \left( 1+ B \arctan\left[ C \left( T-T_c(\mu) \right)  \right] + D \left( T-T_c(\mu) \right) \right).
\eeq
Similarly to the chiral susceptibility, we 
use two possible zero temperature interpolations (6th and 7th order polynomials
of the inverse gauge coupling), two scale
settings, four fit windows. $\kappa$ is obtained either
from a combined $\mu_B^2$ and continuum fit, or separately. 

3) The analysis of the strange susceptibility $\chi^S_2$ goes along the lines of
the chiral condensate. A simplification here is the absence of renormalization.
Since this quantity is the most sensitive to the actual value of strangeness,
before the analysis we correct for the inaccuracies of the strangeness
neutrality condition using the higher $\mu_S$ fluctuations.
Although its inflection point does not have to agree with that of the
chiral condensate, we find that the shifting effect of the chemical potential
is very similar. 

For all three quantities we make a histogram of the results from all 
analyses. For the chiral susceptibility we have two $T>0$ fit forms, two
$T=0$ interpolations, two scale settings, three fit windows and either
separate or combined $\kappa$ extraction and continuum limit. This results
in $2\cdot 2\cdot 2\cdot 3\cdot 2=48$ analyses. For the condensate
we have the same choices but with four fit windows resulting in 64 analyses.
For the strange susceptibility there is no renormalization, thus no $T=0$ interpolation
is needed which leads to 32 analyses.
The central 68\% of the histograms estimates our
systematic error. 
The statistical error is obtained from 1000 bootstrap samples. The two errors
are of similar magnitude and they 
are added in quadrature resulting in our final uncertainties.

We summarize our results for the curvature in Table \ref{tab:kappa}.

\begin{table}[ht]
\begin{center}
\begin{tabular}{l|c}
\hline
 Chiral susceptibility & $0.0158 \pm 0.0013$\\
 Chiral condensate     & $0.0138 \pm 0.0011$ \\
Strange susceptibility & $0.0149 \pm 0.0021$ \\
Susceptibility at $Z=0.4A$ & $0.0149 \pm 0.0017$\\
\hline
\end{tabular}
\caption{\label{tab:kappa}
The curvature ($\kappa$) of the QCD phase diagram in the continuum limit from
various observables. $\kappa$ is fitted in the range $1.2\lesssim\mu_B^I/T\lesssim<2$.
}
\end{center}
\end{table}

The histograms of the three quantities can be joined into a single one leading
to our combined result based on our three observables with strangeness
neutrality:
\begin{equation}
\finalresult\,.
\end{equation}

\begin{figure}[t]
\begin{center}
\includegraphics[width=3in]{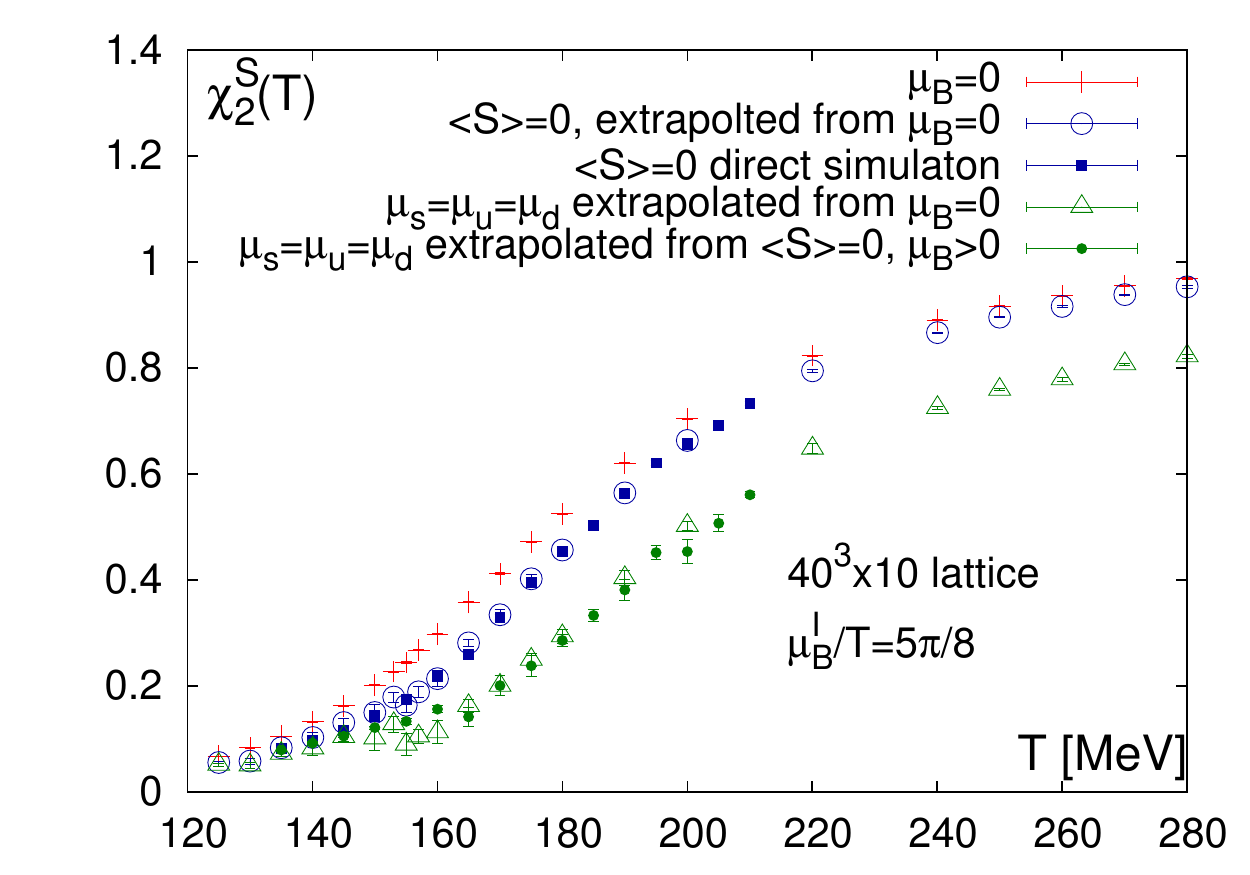}
\end{center}
\caption{\label{fig:extra}
Comparison of the strange susceptibility obtained from direct simulation
and extrapolation for $\mu_B^I/T=5\pi/8$. The blue circles and squares correspond
to the strangeness neutral case obtained via extrapolation from $\mu=0$ and
direct simulations, respectively. The green triangles show the full 
baryo-chemical potential case obtained via extrapolation from $\mu=0$. There are no direct 
simulations in this case but one can extrapolate from the strangeness neutral
direct point (green dots). As a reference the $\mu=0$ data are also shown (red crosses).
}
\end{figure}

We also consider the curvature for the case when not only the
strangeness neutrality, but also proper charge/baryon density ratio is
reproduced (for lead and gold ions: $Z\approx0.4A$). We achieve this by
Taylor-extrapolating the strange susceptibility for every finite 
$\mu_B$ ensemble to leading order, and fitting as before. We conclude
that the difference between $Z=0.4A$ and $Z=0.5A$ phase diagrams is
negligible for small $\mu_B$.

For small enough imaginary chemical potential the analytical and the Taylor
method have to give the same curvature at every lattice resolution.
In the Taylor method one expands the observables in $\mu_B$,
the leading coefficients are calculated from $\mu_B=0$ simulations and then
used to extrapolate to finite $\mu_B$.
Fig.~\ref{fig:extra} shows how this
expansion compares to the direct simulations for our $j=5$ chemical
potential which is the largest one used to extract $\kappa$. A comparison in
the case of full baryo-chemical potential is also shown. 
The extrapolated data are then fitted for $\kappa$ as if they were simulated
at finite $\mu_B^I$. At $N_t=10$ we find $\kappa=0.0131(9)$ from the direct
simulations and $\kappa=0.0115(10)$ from the Taylor expansion.
The agreement indicates that we are still in the linear regime 
and the extraction of $\kappa$ using $j=3,4,5$ is safe. 

Finally we estimate the systematics of the extrapolation to real $\mu_B$.
We include the $j=6$ data points into the analysis and allow for
non-linear $T_c(\mu_B^2)$ fits. We consider fitting $T_c(\mu_B^2)/T_c$ 
with the functions $1+ax$, $1+ax+bx^2$, $(1+ax)/(1+bx)$ and $(1+ax+bx)^{-1}$
with $x=\mu_B^2/T^2$. All these functions are analytic in $x$ and they
represent various analytic continuations of the $T_c(\mu_B^I)$ imaginary $\mu$
phase diagram. The difference between these ansatzes provides a systematic
uncertainty for the real $\mu$ phase diagram.

\begin{figure}[h]
\begin{center}
\includegraphics[width=2.35in]{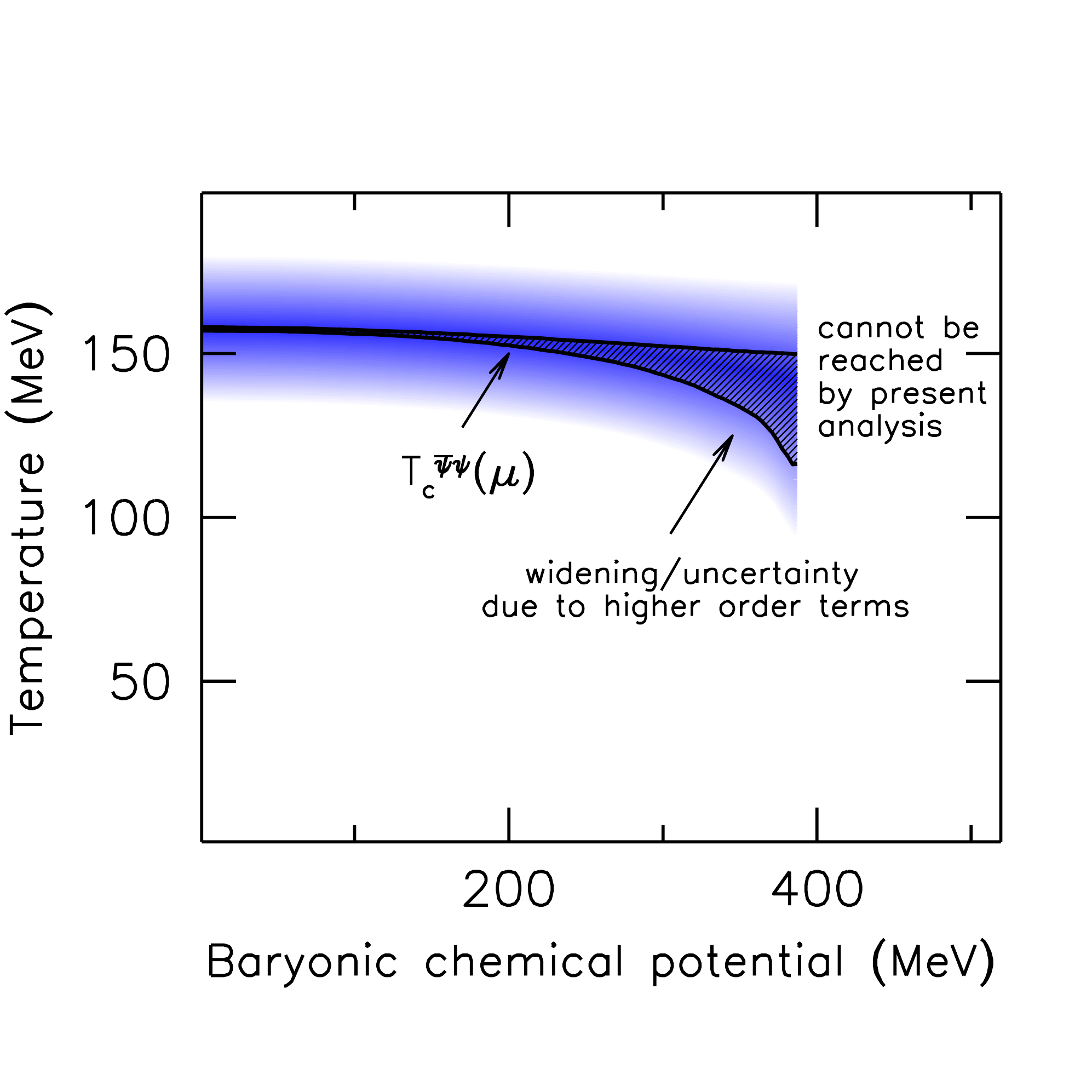}
\includegraphics[width=2.35in]{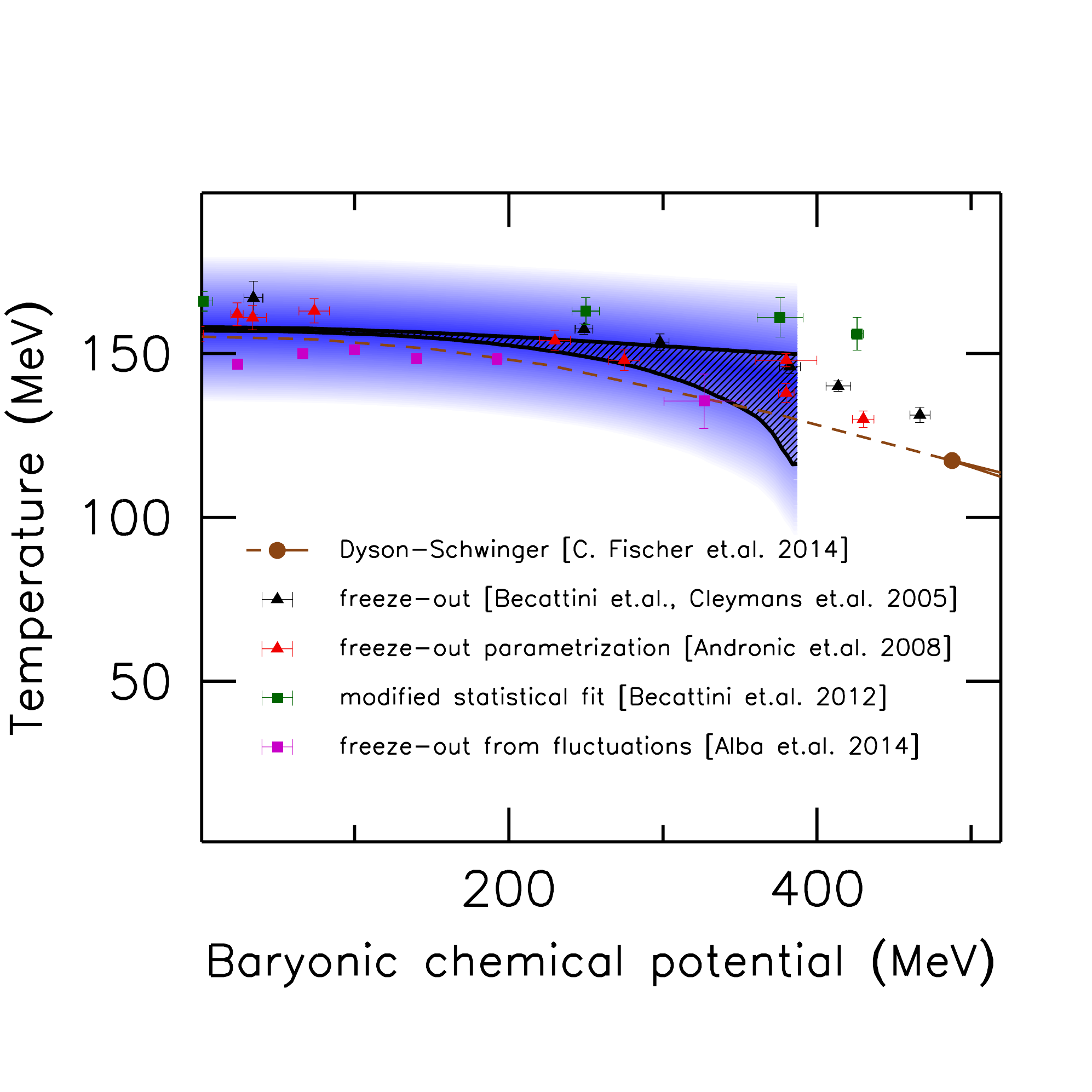}
\end{center}
\caption{\label{fig:phasediagram}
The phase diagram based on the $\mu$-dependent $T_c$ from
the chiral condensate, analytically continued from
imaginary chemical potential. The blue band 
indicates the width of the transition. The shaded black region shows the 
transition line
obtained from the chiral condensate. 
The widening around 300~MeV is coming from the uncertainty of the curvature and from
the contribution of higher order terms, 
thus the application range of the
results is restricted for smaller $\mu$ values. 
For completeness, on the right panel we also show some selected non-lattice results:
the Dyson-Schwinger result of Ref. \cite{Fischer:2014ata} and the 
freeze-out data of
Refs.~\cite{Cleymans:2004pp,Becattini:2005xt,Andronic:2008gu,Becattini:2012xb,Stachel:2013zma,Andronic:2014zha,Alba:2014eba}.  }
\end{figure}

Our main results are depicted in Fig.~\ref{fig:phasediagram}. 
Since the curvature from both the strange susceptibility and from the chiral
condensate/susceptibility are consistent with each other 
we show only one curve.
The curvature from the chiral condensate is our most 
precise result, therefore we 
present the transition line coming from this observable.
The corresponding transition temperature at $\mu$=0 is at 157~MeV.
At intermediate real $\mu_B$ we observe a significant rise in the uncertainty due to the statistical
error on the non-linear $\mu_B^2$-dependence and the ambiguity of the
analytic ansatz. This sets the range of validity for this study.

The present result indicates a stronger curvature than the one presented in Ref.~\cite{Endrodi:2011gv}. There are, however a couple
differences between the definitions/ approaches of the curvature of the present analysis and Ref.~\cite{Endrodi:2011gv}. Note
that the transition is a smooth cross-over, thus different definitions obviously lead to different results.

a. In Ref. \cite{Endrodi:2011gv} we used a vanishing strangeness chemical potential. In the present analysis we use instead vanishing
strange density. The reason for this change is to be as close to the experimental situation as possible. In heavy ion collisions the net
strangeness is zero.

b. It is emphasized in the discussion of Figure 5 of~\cite{Endrodi:2011gv} that only statistical uncertainties were provided. The present
analysis estimates systematic uncertainties coming from various aspects of the analysis as discussed earlier. These are comparable to or in
some cases even larger than the statistical uncertainties. A similar assumption on the systematics of Ref.~\cite{Endrodi:2011gv} would make
the tension between the results much weaker.\hfill\break

\section{Acknowledgments}
The authors thank G. Endrodi for his valuable comments and suggestions.
This project was funded by the DFG grant SFB/TR55. S. D. Katz is funded by
the "Lend\"ulet" program of the Hungarian Academy of Sciences
((LP2012-44/2012).  The work of R.  Bellwied is supported through DOE grant
DEFG02-07ER41521. C. Ratti is supported by the National
Science Foundation through grant number NSF PHY-1513864.
An award of computer time was provided by the INCITE program. This research used resources of the Argonne Leadership Computing Facility, which is a DOE Office of Science User Facility supported under Contract DE-AC02-06CH11357. This research also used resources of
the PRACE Research Infrastructure resource JUQUEEN at FZ-J\"ulich, Germany;
JUQUEEN as large scale project of the Gauss Centre for Supercomputing (GCS);
the QPACE machine supported by the Deutsche Forschungsgesellschaft through
the research program SFB TR55; and the GPU cluster at the Wuppertal University.

\bibliography{thermo}{}
\end{document}